## Flickers Forecasting In CRT Using Stochastic Analysis

A. A. KHAN, S. SOOMRO*, A. G. MEMON**

Institute of Business Management, Pakistan
Email: safeeullah.soomro@biztekian.com,:ghafoor@usindh.edu.pk

**Abstract**: Videos are composed of sequence of interrelated frames. There is a minute difference among frames. Flicker is an error which is found in every video. It is like a checker box in a video, there are several reasons behind flickers generation, one of the main reasons is refresh rate of the monitor and second reason is number of frames per second in a video. The main objective of this study is to propose and develop a framework that identifies flicker location and minimizes the flickers rate. Analysis shows that flickers can be minimize by adjusting the persistence of pixel and higher refresh rate of CRT monitor. Further we have compared different isotopes of phosphorous pixels and generate its graphs. This paper highlighted the cause of flicker and its avoidance .Statistical research proves that proposed algorithm improves the video quality and reduce flickers ratio up to 90%.

### 1. INTRODUCTION

Series of individual images in a movie clip called frames. When these images are shown rapidly in succession, a viewer has the illusion that motion is occurring. The viewer cannot observe the flickering in a movie clip due frame continuation is known as persistence of vision. It is due to eye retains on a visual image for a fraction of a second after the source has been removed. While watching a movie observer noticed unwanted checker boxes when a movie scene changed this can be called as Flickers. These flickers are due to abrupt variation of intensities in the phosphorous pixels Yoon, (2011). Flicker forecasting and avoidance can improve the video quality and provides flicker free environment. It is difficult to calculate abruption in pixels but can be forecast by stochastic method. This process contains family of random variables which is indexed by the parameter "t" and defined on a common probability space ($\Omega$, F, P) whereas t, w , $\Omega$, $X_t$ (w) is a random variable. For each w, T, $X_t$ (w) is called a sample function or realization of the process. The result shows that intensity barrier is one of the main causes behind this error, by comparing two pixels of two separate frames shows the intensity gap. The intensity gap among two pixels can easily determine by probability of selected pixels. Hence by comparing the pixels intensities it is easy to calculate flickers. The analysis shows the flicker will emerge in a clip if probability among pixels is above 0.05.

In order to reduce flicker we add a new pixel among higher probability pixels. This additional pixel contains the mean pixel values of consecutive pixels that can reduce the luminosity of a flicker. On fast motion scenes, a variable bit rate uses more bits than it does on slow motion scenes of similar duration yet achieves a consistent visual quality. High definition video transmission, low bandwidth can also produce flicker or checkers. Human Eye cannot notice it due to persistence of vision and beta movement (Farrell, *et al.,* 1987)

On the other hand, the Flat-panel liquid crystal display (LCD) monitors do not suffer from flicker even if its refresh rate is 60 Hz or even lower. This is because LCD pixels open to allow a continuous stream of light to pass through until instructed by the video signal to produce a darker color. In contrary monitors are composed of different types of phosphorous isotopes mainly P31, BARCO D65 P4, DP104 or RGB P22 etc. Further by comparing these pixel's types it will be easy to get the best flicker regression element. In other words it can be stated as flicker rate is equal to flicker regression of the phosphorous pixels with pixel luminance energy decay time (Najim, *et al.,* 2006).

++ Corresponding author: email: Adnan_hiit@yahoo.com,
*Institute of Business and Technology, BIZTEK, Karachi, Pakistan
**Institute of Mathematics and Computer Science, University of Sindh, Pakistan

$$\text{Flicker rate} = \frac{\text{Flicker regression of the phosphorous pixels}}{\text{Pixel luminance decay time}}$$

## 2. MATERIAL AND METHODS
**System Design Platform:**

Our aim is to decrease screen luminance energy until perceived flicker disappeared then increase screen luminance energy until perceived flicker reappears.

```
INITIALIZE:
LOAD: Import a scene
SET: Frame[a]=0;
LOOP: If F[a] ≤ F[n], F[a] ;then Increment F[b] = F[a]+1
COMPARE: F[a] with F[b]
CONVERSION: F[a]_Image → F[a]_Matrix ; F[b]_Image → F[b]_Matrix
CALCULATE: Probability:= Prob{F[a] (m*n)_i, F[b](m*n)_i}  ∴ i=pixel location
LOOP: Probability ≥ 0.05
SPLIT F[a]i_Color → F[a]i_Red || F[a]i_Green || F[a]i_Blue
SPLIT F[b]i_Color → F[b]i_Red || F[b]i_Green || F[b]i_Blue
Mean (F[a], F[b])i_Red, Mean (F[a], F[b])i_Green, Mean (F[a], F[b])i_Blue.
New Pixel_color := Add (Mean (F[a], F[b])_Red, Mean (F[a], F[b])_Green, Mean (F[a], F[b])_Blue.)
INSERT (New Pixel_color) := (F[a]_i, F[b]_i)
END
END
```

**Fig. 1. Proposed Algorithum**

There are two main loops in this algorithm one is for increment in frames and second one is for probability calculation (see Table 3 to 8). Software determines the need of mean pixel insertion. It segregates image into three colors then calculate mean value of pixels and insert it if it is required.

**1. The System Design Tools:**

Noise is a universal entity which is found everywhere it can easily been observed in videos as well. Unfortunately when flicker or unwanted noise appears between two frames, it is because of abrupt changes in luminance; refresh frequencies, screen luminance in terms of wave length, refresh rate of screen phosphors, and the difference between viewer and Cathode Ray Tube CRT. Consequently flicker emerges in a film; in this context we have calculated the probability of two pixels of the two different frames. Hence research shows that if the probability is greater than 0.05 than flicker emerges. Therefore proposed algorithm uses stochastic theorem to forecast the flicker and thus can reduce the flicker luminosity. Software calculates the pixels intensities and updates the flicker pixel with it. There are total 8 different images or 4 different pairs will be used for this experiment. The name of software is "flicker" which is developed in Matlab version seven releases 14, this "m" file calculates flickers and show results.

Again, the average of the two luminance thresholds was used, together with the refresh rate and phosphorous time constant. Further software compares the types of phosphorous pixels and generates graphs. These graphs depict which type is suitable for flicker free monitors. An algorithm has been developed to solve this problem. This Algorithm works in two ways one is identification and second one reduction. Proposed algorithm is as follows.

### 1.1 Flickers reduction Electronically

Image in a video contains series of luminance variation with exponentially decaying persistence of any pixel. If we can compute the amplitude coefficient of the fundamental frequency of the time-varying screen luminance (as per Oppenhein & Willsky, 1983) has mentioned. Amplitude coefficient is the formula that can compare the existing Phosphorous pixels. Formula of amplitude coefficients is as follows.

$AmpCoeff(F) = \frac{2}{[1+(\alpha\omega)^2]^{1/2}}$ .where as α is the decaying or persisting time of a phosphor pixel 'ω' is "2Πf" and "f" is the refresh rate of any monitor. Willsky further explain the degree of visual of angles which is $2[tan^{-1}\left(\frac{D}{2V}\right)]$ , whereas D is related with the display area measured in millimeters and V = 1.5 feet or 500 mm. Forth coming graphs provides a good summarized version of this paper (Donovan 2011)

## 1.2 The Flickers identification

White light is composed of seven colors. It can be segregated into seven different colors. There are total eight colors in Matlab programming list is as follows. [6] as shown in **(Table 1).**

**Table 1. Example of color computerized codes**

| Color Name | Symbol | Computer Code |
|---|---|---|
| Yellow | y | [1 1 0] |
| Magenta | m | [1 0 1] |
| Cyan | c | [0 1 1] |
| Red | r | [1 0 0] |
| Green | g | [0 1 0] |
| Blue | b | [0 0 1] |
| White | w | [1 1 1] |
| Black | k | [0 0 0] |

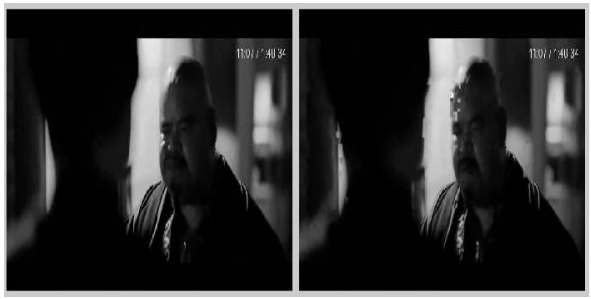

**Fig.2. Input frames.**

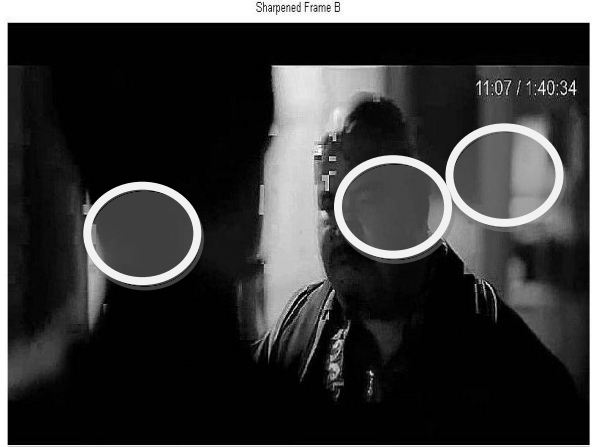

**Fig.3. The Flickers demonstration.**

There are two frames from a movie that contains flickers (**see Fig. 2).** After image enhancement flickers are more visible to the human eye (**see Fig.3 and 4**).

## 2. The Development of Proposed Framework:

Matlab is the main tool for programming and simulations. The following is the data flow diagram that elaborates steps by step working.

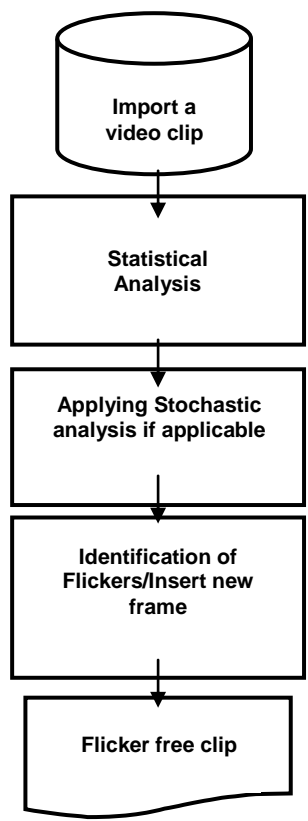

**Fig. 4. Data flow diagram of proposed frame work**

This framework helps in the identification process of flickers and its reduction. Software imports a video clip; convert each frame into separate red, green, blue signals. It determines the probability and test results. If probability of selected pixel value is above 0.05 it updates it with new one.

## 1.3 The Statistical Model:

Software identifies those pixels whose probability value is greater than 0.05. A real scenario is considered for step-wise development of proposed frame work. This real case information is presented in following (**see Fig. 5 to 11**).

| Actual | Relation between colors(F[a] with F[b]) | | | | | | | |
|---|---|---|---|---|---|---|---|---|
| | Black | Blue | Green | Cyan | Red | Magenta | Yellow | White |
| Black | 0 | 0.5 | 1 | 1.5 | 2 | 2.5 | 3 | 3.5 |
| Blue | 0.5 | 0 | 1.5 | 2 | 2.5 | 3 | 3.5 | 4 |
| Green | 1 | 1.5 | 0 | 2.5 | 3 | 3.5 | 4 | 4.5 |
| Cyan | 1.5 | 2 | 2.5 | 0 | 3.5 | 4 | 4.5 | 5 |
| Red | 2 | 2.5 | 3 | 3.5 | 0 | 4.5 | 5 | 5.5 |
| Magenta | 2.5 | 3 | 3.5 | 4 | 4.5 | 0 | 5.5 | 6 |
| Yellow | 3 | 3.5 | 4 | 4.5 | 5 | 5.5 | 0 | 6.5 |
| White | 3.5 | 4 | 4.5 | 5 | 5.5 | 6 | 6.5 | 0 |

**Fig.5. Relationship among pixels .**

| Stockestic Matrix(Colom wise) | | | | | Probability see paper | | | |
|---|---|---|---|---|---|---|---|---|
| | Black | Blue | Green | Cyan | Red | Magenta | Yellow | White |
| Black | 0 | 0.029412 | 0.05 | 0.065217 | 0.076923 | 0.086207 | 0.09375 | 0.1 |
| Blue | 0.035714 | 0 | 0.075 | 0.086957 | 0.096154 | 0.103448 | 0.109375 | 0.114286 |
| Green | 0.071429 | 0.088235 | 0 | 0.108696 | 0.115385 | 0.12069 | 0.125 | 0.128571 |
| Cyan | 0.107143 | 0.117647 | 0.125 | 0 | 0.134615 | 0.137931 | 0.140625 | 0.142857 |
| Red | 0.142857 | 0.147059 | 0.15 | 0.152174 | 0 | 0.155172 | 0.15625 | 0.157143 |
| Magenta | 0.178571 | 0.176471 | 0.175 | 0.173913 | 0.173077 | 0 | 0.171875 | 0.171429 |
| Yellow | 0.214286 | 0.205882 | 0.2 | 0.195652 | 0.192308 | 0.189655 | 0 | 0.185714 |
| White | 0.25 | 0.235294 | 0.225 | 0.217391 | 0.211538 | 0.206897 | 0.203125 | 0 |
| Sum | 1 | 1 | 1 | 1 | 1 | 1 | 1 | 1 |
| Mean | 0.125 | 0.125 | 0.125 | 0.125 | 0.125 | 0.125 | 0.125 | 0.125 |
| Variance | 0.006696 | 0.006001 | 0.005313 | 0.004696 | 0.004161 | 0.003697 | 0.003296 | 0.002946 |
| S. Dev | 0.081832 | 0.077468 | 0.072887 | 0.06853 | 0.064502 | 0.060805 | 0.05741 | 0.054281 |

**Fig.6. The Stochastic Matrix column wise.**

| Z- Values | | | | | | | | |
|---|---|---|---|---|---|---|---|---|
| | Black | Blue | Green | Cyan | Red | Magenta | Yellow | White |
| Black | -1.52753 | -1.23391 | -1.02899 | -0.87236 | -0.74536 | -0.63799 | -0.54433 | -0.46057 |
| Blue | -1.09109 | -1.61357 | -0.68599 | -0.55514 | -0.44721 | -0.35444 | -0.27217 | -0.19739 |
| Green | -0.65465 | -0.47458 | -1.71499 | -0.23792 | -0.14907 | -0.07089 | 0 | 0.065795 |
| Cyan | -0.21822 | -0.09492 | 1.9E-16 | -1.82402 | 0.149071 | 0.212664 | 0.272166 | 0.328976 |
| Red | 0.218218 | 0.284747 | 0.342997 | 0.396526 | -1.93793 | 0.496217 | 0.544331 | 0.592157 |
| Magenta | 0.654654 | 0.664411 | 0.685994 | 0.713746 | 0.745356 | -2.05576 | 0.816497 | 0.855337 |
| Yellow | 1.091089 | 1.044074 | 1.028992 | 1.030967 | 1.043498 | 1.063322 | -2.17732 | 1.118518 |
| White | 1.527525 | 1.423737 | 1.371989 | 1.348188 | 1.341641 | 1.346874 | 1.360828 | -2.30283 |
| Sum | 0 | 0 | 0 | 0 | 2E-15 | 0 | 0 | 0 |
| Mean | 0 | 0 | 0 | 0 | 2.5E-16 | 0 | 0 | 0 |
| Variance | 1 | 1 | 1 | 1 | 1 | 1 | 1 | 1 |
| S. Dev | 1 | 1 | 1 | 1 | 1 | 1 | 1 | 1 |

**Fig.7. The 'Z' values**

| Z- Values (Probabilities) | Black | Blue | Green | Cyan | Red | Magenta | Yellow | White |
|---|---|---|---|---|---|---|---|---|
| Black | 0.0643 | 0.1093 | 0.1539 | 0.1922 | 0.2266 | 0.2611 | 0.2946 | 0.3228 |
| Blue | 0.1379 | 0.0537 | 0.2483 | 0.2877 | 0.33 | 0.3632 | 0.3936 | 0.4247 |
| Green | 0.2587 | 0.3192 | 0.0436 | 0.4052 | 0.1251 | 0.242 | 0.5 | 0.242 |
| Cyan | 0.0146 | 0.4641 | 0.5 | 0.0359 | 0.5596 | 0.5832 | 0.6064 | 0.6293 |
| Red | 0.5832 | 0.6103 | 0.6331 | 0.6517 | 0.0262 | 6915 | 0.7054 | 0.7224 |
| Magenta | 0.7422 | 0.7454 | 0.7549 | 0.7611 | 0.7704 | 0.0197 | 0.7939 | 0.8051 |
| Yellow | 0.8621 | 0.8508 | 0.8485 | 0.8485 | 0.8508 | 0.8554 | 0.0164 | 0.8686 |
| White | 0.937 | 0.9222 | 0.9147 | 0.937 | 0.9222 | 0.9115 | 0.9131 | 0.0107 |

**Fig.8. The 'Z' values Probabilities**

Stockestic Matrix(Row wise)

| | Black | Blue | Green | Cyan | Red | Magenta | Yellow | White | Sum | Mean | Var. | S.Dev. |
|---|---|---|---|---|---|---|---|---|---|---|---|---|
| Black | 0 | 0.029412 | 0.05 | 0.065217 | 0.076923 | 0.086207 | 0.09375 | 0.1 | 0.501509 | 0.062689 | 0.00104 | 0.032244 |
| Blue | 0.035714 | 0 | 0.075 | 0.086957 | 0.096154 | 0.103448 | 0.109375 | 0.114286 | 0.620934 | 0.077617 | 0.001405 | 0.037481 |
| Green | 0.071429 | 0.088235 | 0 | 0.108696 | 0.115385 | 0.12069 | 0.125 | 0.128571 | 0.758005 | 0.094751 | 0.001614 | 0.040181 |
| Cyan | 0.107143 | 0.117647 | 0.125 | 0 | 0.134615 | 0.137931 | 0.140625 | 0.142857 | 0.905818 | 0.113227 | 0.001964 | 0.044317 |
| Red | 0.142857 | 0.147059 | 0.15 | 0.152174 | 0 | 0.155172 | 0.15625 | 0.157143 | 1.060655 | 0.132582 | 0.002532 | 0.050317 |
| Magenta | 0.178571 | 0.176471 | 0.175 | 0.173913 | 0.173077 | 0 | 0.171875 | 0.171429 | 1.220336 | 0.152542 | 0.003329 | 0.057698 |
| Yellow | 0.214286 | 0.205882 | 0.2 | 0.195652 | 0.192308 | 0.189655 | 0 | 0.185714 | 1.383497 | 0.172937 | 0.004346 | 0.065925 |
| White | 0.25 | 0.235294 | 0.225 | 0.217391 | 0.211538 | 0.206897 | 0.203125 | 0 | 1.549245 | 0.193656 | 0.005568 | 0.074617 |

**Fig.9. The Stochastic Matrix Row wise**

Stockestic Matrix(Row wise) [Z-Values]- Probabilites

| | Black | Blue | Green | Cyan | Red | Magenta | Yellow | White |
|---|---|---|---|---|---|---|---|---|
| Black | 0.0262 | 0.1515 | 0.3483 | 0.5279 | 0.67 | 0.7673 | 0.8315 | 0.8749 |
| Blue | 0.1314 | 0.0192 | 0.242 | 0.59887 | 0.6879 | 0.7549 | 0.8023 | 0.8365 |
| Green | 0.281 | 0.4364 | 0.0091 | 0.6331 | 0.695 | 0.7389 | 0.7734 | 0.7995 |
| Cyan | 0.4443 | 0.5359 | 0.6026 | 0.0054 | 0.6844 | 0.7123 | 0.7291 | 0.7486 |
| Red | 0.5793 | 0.6103 | 0.6368 | 0.648 | 0.0043 | 0.6736 | 0.6808 | 0.6879 |
| Magenta | 0.6736 | 0.6591 | 0.6517 | 0.6443 | 0.6406 | 0.0041 | 0.6331 | 0.6293 |
| Yellow | 0.7357 | 0.6915 | 0.6591 | 0.6331 | 0.6141 | 0.4013 | 0.0044 | 0.5753 |
| White | 0.7764 | 0.7088 | 0.6628 | 0.6255 | 0.5948 | 0.5714 | 0.5517 | 0.0047 |

**Fig.10. The Stochastic Matrix column wise Probabilities**

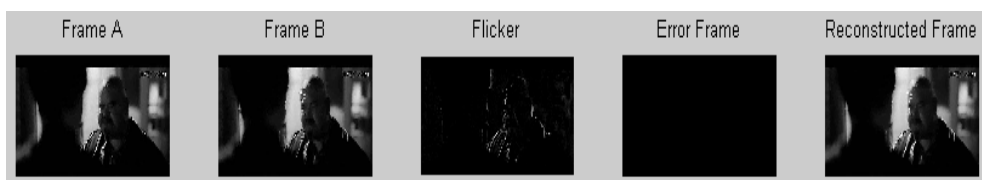

**Fig.11.Working of Algorithm**

Software import video clip , identifies flickers by calculating probability, generate error frame and develop reconstructed image. The proposed algorithm develops a new frame called reconstructed frame that will be inserted between two frames and produces flicker free video clip. The following result demonstrates flickers from a scene and related percentage **(See Fig.)**.

**1.4 Electronically Evaluation of phosphorous pixels:**

The forth coming result indicates the possibility of flicker prediction in any display terminal. Proposed algorithm proves the possibility flicker-free video. Phosphorous pixel P31, DP104 provides 90% luminance persistence. The flicker-free predictions will vary as a function of display phosphor persistence; refresh frequency, luminance, and display size *www.wikipedia.com/stochastic*

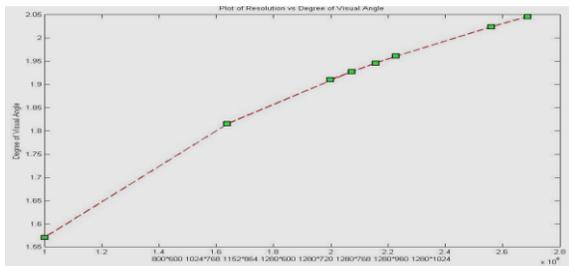

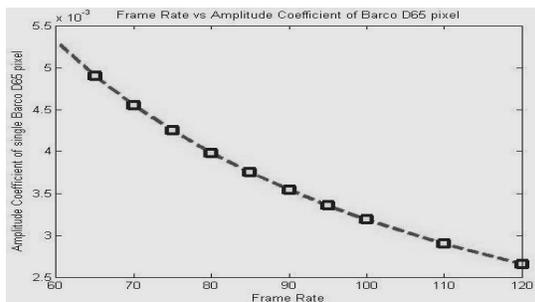

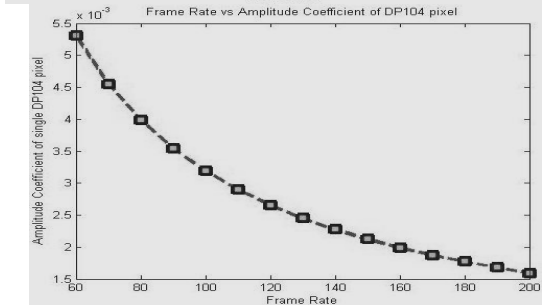

**Graph 1 shows the relationship between monitor resolution and degree of visual angle. The visual angle of an image is increased by enhancing the monitor resolution. Graph2 and 3 shows the relationship between amplitude coefficient of D65, P31 and Dp104 with its refresh rate. You can see that DP104or graph 3 provides the better result among all. It means it is clear that it got higher refresh rate and less phosphorous persistence.**

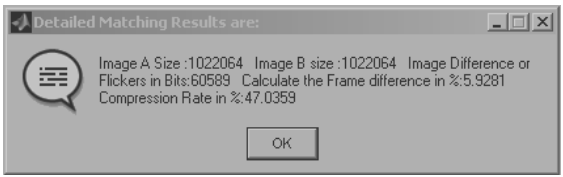

**Fig.12.Amplitude Coefficient of P4, D65 and DP 104**

**5. CONCLUSION**

Monitor is composed of fast displaying phosphorus. It has relatively large amplitude coefficients which require less energy or relatively less screen luminance to be seen. This problem can be addressed by two methods one is stochastic theorem and second is Eriksson and Backstrom method of predicting flickers. The fast displays phosphors DP104 appears to be a less flicker at low luminance due to its short decay constant. The screen luminance for perceived flicker is proportional to the decay constant of the phosphor. Flicker prediction lies in stimulus energy which is the fundamental frequency of the temporally modulated screen luminance. Second method is related with stochastic theorem for flicker prediction in a movie. If a movie clip contains undesired noise which reflects the pixel transition between two frames are not smooth. The size of the image A is 998.10 K Byte and the size of the image B is 998.10 K Byte which is shown in fig.5. Here two different frames A and frame B are used (fig 2) then algorithm identifies flickers using stochastic model and shows almost 6% of flickers are there in a scene. The reconstructed frame D is an outcome of initial image A and the flicker frame (the image D) see Figure 4. Hence it reduces the flicker and provides smooth frame transition for high definition videos. Despite of flicker reduction this algorithm compresses frames and provides a new dimension for low bandwidth HD video transmission.